\documentclass[aps,prl,twocolumn,nofootinbib]{revtex4-1}
\usepackage{bm}
\usepackage{latexsym}
\usepackage{amsmath,amsfonts,amssymb}
\usepackage{graphicx,epsfig}
\usepackage[utf8]{inputenc}
\usepackage{array}
\usepackage{wrapfig}
\usepackage{multirow}
\usepackage{tabularx}
\usepackage{psfrag}
\usepackage{subfigure}
\usepackage{amsthm}
\usepackage{color}
\usepackage[dvipsnames]{xcolor}

\def\be{\begin{equation}}
\def\ee{\end{equation}}
\def\ber{\begin{eqnarray}}
\def\eer{\end{eqnarray}}

\usepackage{amsmath}
\usepackage{braket}
\usepackage[normalem]{ulem}
\useunder{\uline}{\ul}{}

\begin{document}
\title{New insights on the quantum-classical division in light of Collapse Models}

\author{Fernanda Torres}
\email{mftorrescabrera@uh.edu}
\affiliation{Department of Physics, University of Houston, Houston, Texas 77024-5005, USA}
\affiliation{Facultad de Ciencias - CUICBAS, Universidad de Colima, Colima, C.P. 28045, M\'exico}

\author{Sujoy K. Modak}
\email{smodak@ucol.mx}
\affiliation{Facultad de Ciencias - CUICBAS, Universidad de Colima, Colima, C.P. 28045, M\'exico}

\author{Alfredo Aranda}
\email{fefo@ucol.mx}
\affiliation{Facultad de Ciencias - CUICBAS, Universidad de Colima, Colima, C.P. 28045, M\'exico}

\begin{abstract}
We argue, in light of Collapse Model interpretation of quantum theory, that the fundamental division between the quantum and classical behaviors might be analogous to the division of thermodynamic phases. A specific relationship between the collapse parameter $(\lambda)$ and the collapse length scale ($r_C$)  plays the role of the coexistence curve in usual thermodynamic phase diagrams. We further claim that our functional relationship between  $\lambda$ and $r_C$ is strongly supported by the existing International Germanium Experiment (IGEX) collaboration data. This result is preceded by a brief discussion of quantum measurement theory and the Ghirardi-Rimini-Weber (GRW) model applied to the free wavepacket dynamics.

\end{abstract}
\maketitle

Quantum mechanics (QM) persists in providing opportunities for interpretations and understanding. Among the questions that keep us interested are the ``measurement problem'' and the issue of the inter-phase between classical and quantum phenomena, i.e. when could one say firmly that a distinction between the two occur?, where is the boundary? Although as old as QM itself, these interesting questions remain subject to investigation and can lead to (have led to) interesting ideas. 
In this letter we explore a particular setting -- within the realm of so-called Collapse Models \cite{pearle}-\cite{Bassi:2012bg} -- and make a case that identifying and classifying quantum and classical ``phases'' might be similar to the identification and classification of phases in the thermodynamic/magnetic/QGP systems.

``Measurement problem'' is explicit when we treat an apparatus quantum mechanically to explain the outcome of a measurement. We inevitably land in an unbreakable von-Neumann chain (VNC) for which Collapse Models \cite{pearle}-\cite{Bassi:2012bg} come to the rescue. To see that consider $A$ to be the apparatus that measures a system $S$ having an observable ${\cal O}$. Then, the time evolution of the system-apparatus is controlled by a linear unitary operator $U$, whose evolution conserves the conjugation relations, and is given by, 
\begin{equation} 
\label{eqn:U}
    U|\psi_0 \rangle |A_0 \rangle = \sum_n c_n |\psi_n \rangle |A_n \rangle.
\end{equation}
The initial state of the system is $|\psi_0\rangle$ and of the apparatus is $|A_0\rangle$. Time evolution makes a linear superposition of the $S+A$ system due to the use of the unitary operator $U$ which, in turn, invites some troubles. First the eventual appearance of a statistical mixture of states originating from the system and the apparatus after the measurement is performed. But even to get to the statistical mixture invites a second problem: the information about the entire system ($S+A$) during the measurement process is not complete. It is necessary to have another apparatus $B$ that can measure the final state of the apparatus $A$ in order to know the final state of $S+A$. In this way, cumulatively, we end up with a measurement process in a VNC where finally some conscious mind (observer) is needed to determine a stopping point to break it. Thus observer becomes an inseparable part of a physical theory!

In the well-known von-Neumann model of quantum measurement, treating the system $S$ and apparatus $A$ quantum mechanically, initially one associates $S$ in a  state $|\psi_{0}\rangle$  and $A$ in a   well-localized state $|A_0\rangle$. Without  loss of generality, one can set the pointer's position  (signalling the center of mass position of $A$) at zero and assuming $A$ is static its momentum is also zero. Initially there is no interaction between the system and the apparatus. The complete system is $S+A$, and the total Hamiltonian is given by $H = H_A + H_{int}$
where $H_A = P^2/2M$ refers to the Hamiltonian of the apparatus of mass $M$ when treated as a free particle. To do so, the pointer mass is made large enough to make the pointer wavepacket localized for a time much longer than the duration of measurement. Also, there is an interaction between the apparatus and the system given by $H_{int}$. Due to the smallness of the quantum system we can neglect the free Hamiltonian of the system $S$ for simplicity.  The interaction between the $S$ and $A$ can be modelled in the following way \cite{87}:  
\begin{equation}
    H_{int} = \frac{d\beta}{dt} f({\cal{O}}) P
\end{equation}
where $d\beta/dt$ is the change of the interaction during the interval $(t_0,t_1)$. $\beta(t)$ vanishes for $t \le t_0$, then increases from $t_0$ to $t_1$, and at some later time $t \ge t_1$ becomes one. Thus the interaction Hamiltonian is zero before $t_0$ and after $t_1$. In this model details of the interaction between the system and the apparatus are unspecified but the overall consequence of such a process is contemplated. The operator $f({\cal{O}})$ is a function of the observable to be measured which, from a dimensional analysis,  has a dimension of length.  Once an interaction begins, the $S+A$ system evolves in the following manner:
\begin{equation}
\label{sup}
\begin{split}
    |\Phi(t)\rangle & = e^{-\frac{i}{\hbar}\int H dt} |\psi_0\rangle |A_0\rangle 
    \\
    %& =
    %e^{-\frac{i}{\hbar}H_A t} e^{-\frac{i}{\hbar}\beta(t)f({\cal O})P} |\psi\rangle |A_0\rangle 
    %\\
    & = 
    \sum_n c_n e^{-\frac{i}{\hbar}H_A t} e^{-\frac{i}{\hbar}\beta(t)f(O_n)P} |\psi_n\rangle |A_0\rangle 
    \\
    & =
    \sum_n c_n |\psi_n\rangle |A_n(t)\rangle. %\label{sup}
\end{split}
\end{equation}
In the above equation  $|\psi_n\rangle$ and $|A_n\rangle$ are the final states of the system and the pointer. This equation is truly applicable for the interval $[t_0,t_1]$ and it indicates that the system and apparatus are in an entangled state; that is, the eigenstate $| \psi_n \rangle$ of the system is entangled with the eigenstate of the apparatus, given by
\begin{equation}
\label{ans}
    |A_n(t)\rangle = e^{-\frac{i}{\hbar}H_A t} e^{-\frac{i}{\hbar}\beta(t)f(O_n)P} |A_0\rangle.
\end{equation}
Therefore the whole $S+A$ system is now in a superposition of states and due to entanglement when, in the apparatus space, the pointer picks an eigenstate $| A_n \rangle$, the above superposition in \eqref{sup} ceases to exist making the system collapse to its eigenstate $| \psi_n \rangle$. 
A collective explanation in terms of the density matrix is the following: before the wavepacket reduction the statistical operator is
\begin{equation}
    \rho = \sum_{m,n} c_n c_m^* |\psi_m\rangle \langle \psi_n|\otimes |A_m\rangle \langle A_n|
\end{equation}
and when the measurement is performed the pointer state $\langle Q \rangle$ is found near a specific $f({\cal O}_n)$ which  implies that the apparatus state is collapsed to $|A_n\rangle$  and thus the density operator above will be diagonal
\begin{equation}
    \rho = \sum_{n} |c_n|^2 |\psi_n\rangle \langle \psi_n|\otimes |A_n\rangle \langle A_n|.
\end{equation}
When such a diagonalization takes place we interpret this as the reduction of the wavepacket and, in doing so, the initial pure density operator evolves into a mixed state \cite{Esp} as required by a measurement.

%\begin{figure}[t!]
%\centering
%\includegraphics[width=0.5\textwidth]{Q-measure.png}
%\caption[]{Schematic diagram describing the idea of quantum theory of measurement, following  Benatti, Ghirardi, Rimini and Weber \cite{87},  considering both the system $S$ and apparatus $A$ quantum mechanically. For details see text. }
%\label{measure}
%\end{figure}

Collapse model interpretation of quantum mechanics \cite{pearle}-\cite{Bassi:2012bg}  has an additional feature of stochasticity along with the linear Schr\"odinger evolution, working in a way that microscopic quantum behavior is practically preserved, but importantly, macroscopic superpositions are effectively suppressed assuring that the macroscopic wave functions are well localized in space without needing an observer making a measurement. The Ghirardi-Rimini-Weber (GRW) version \cite{grw} considers ``measurement like effects'' happen discretely in time following a Poisson distribution with or without measurement. A parameter $\lambda$ is introduced as the mean frequency of such discrete collapse events. Once again, the reduction of the wavepacket is represented by the diagonalization of the density operator, which when collapse happens suffers a jump $\rho \rightarrow T_q[\rho]$ in the vicinity of $q$ \cite{grw,87} such that
\begin{eqnarray}
  T_q[\rho] = \sqrt{\frac{\alpha}{\pi}} \int_{\mathrm{R}}{dx} e^{-\frac{\alpha}{2}(q-x)^2} ~\rho~ e^{-\frac{\alpha}{2}(q-x)^2},
  \label{new1}
\end{eqnarray}
and $\alpha$ provides the second important parameter -- the collapse length scale $r_C=1/{\sqrt{\alpha}}$. The length $r_C$ has the following role -- if initially two given states $|\psi_1\rangle$ and $|\psi_2\rangle$ (making the density operator) are individually localized in region less than $r_C$ while the distance between them is more than $r_C$, we consider them distinct states and truly in superposition. The density matrix constructed by them is initially pure and satisfies the modified Schr\"odinger evolution,
\begin{equation}
    \frac{d\rho}{dt} = -\frac{i}{\hbar} [H, \rho] + \lambda (T_q[\rho] - \rho). \label{cr1}
\end{equation}
By virtue of the above equation it takes a time equivalent to $1/\lambda$ for the initial pure density operator to become diagonal or mixed. Note that, the above equation \eqref{cr1}  can be easily generalized for $N$ particle systems by scaling $\lambda$ to $N\lambda$ \cite{grw,87}.

For an ensemble of free wavepacket systems \eqref{cr1} becomes
\begin{equation}
\label{eqn:48}
\begin{split}
    \frac{\partial}{\partial t} \langle q'|\rho(t)|q'' \rangle = \frac{i \hbar}{2m} \left(\frac{\partial^2}{\partial q'^2} - \frac{\partial^2}{\partial q''^2} \right) \langle q'|\rho(t) |q'' \rangle
    \\
    -\lambda (1-e^{-(\alpha/4)(q'-q'')^2}) \langle q'|\rho(t)|q''\rangle.
\end{split}
\end{equation} 
The above differential equation can be rigorously solved \cite{grw,87} to yield
%\begin{widetext}
\begin{eqnarray}
%\label{eqn:49}
    \langle q'| \rho(t) |q''\rangle &=& \frac{1}{2\pi \hbar} \int_{-\infty}^{\infty} dk \int_{-\infty}^{\infty} dy  ~e^{-(i/\hbar)k y} \nonumber \\
    && \times F(\lambda, k,q'-q'',t) \langle q'+y| \rho_S(t) |q''+y \rangle \nonumber,
\end{eqnarray}
%\end{widetext}
where
\begin{equation}
\label{ffn}
    F(\lambda, k,q,t) = e^{-\lambda t \left(1-\frac{1}{t} \int_0^t d \tau ~e^{-(\alpha/4)(q-k\tau/m)^2} \right)}
\end{equation}
is the term driving the nonlinear dynamics of state reduction and the  $\rho_S(t)$ is defined in the Schr\"odinger picture. At an initial time $t=0$, $F(\lambda, k,q,t)=1$ and there is no distinction from the standard Schr\"odinger initial value. However, at any later time we have a modified dynamics which is influenced by the occasional collapse processes for the ensemble of identically prepared initial free wavepackets. Following \cite{grw} one can calculate expectation values of various observables such as position, momentum and their functions by derivating the function defined in \eqref{ffn}. Up to the quadratic order of operator expectation values, one obtains \cite{grw} 
\begin{align}
     & \langle \hat{q} \rangle_{GRW} = \langle \hat{q} \rangle_S \label{si}\\
     & \langle \hat{q}^2 \rangle_{GRW} =  \langle \hat{q}^2 \rangle_S + \frac{\alpha \lambda \hbar^2}{6m^2} t^3\\
     & \langle \hat{p} \rangle_{GRW} = \langle \hat{p} \rangle_S \\
    &  \langle \hat{p}^2 \rangle_{GRW} =  \langle \hat{p}^2 \rangle_S + \frac{\alpha\lambda \hbar^2}{2} t \\
    &   \langle \hat{q} \hat{p} \rangle_{GRW} = \langle \hat{q} \hat{p} \rangle _{S} + \frac{\alpha \lambda \hbar^2}{4m} t^2 \label{sf}.
    \end{align}

In the remaining part of this work we show remarkable usefulness of above results not only for constraining Collapse Models  but also for the general understanding on the quantum-classical division. 

First recall that for an ensemble of identically prepared free wavepackets one can calculated average rate of expansion, using Ehrenfests's equations, due to the uncertainty principle in the standard Schr\"odinger picture (without any collapse mechanism). The square of the average width $\langle \xi_{S} \rangle = \langle \Delta q_S^2 \rangle = \langle q^2 \rangle_S - \langle q\rangle_S^2$ satisfies the following second order equation \cite{messiah},
\begin{equation}
     \frac{d^2}{dt^2}{\langle \xi \rangle } = \frac{2\langle \varpi_0 \rangle}{m^2},
\end{equation}
where $\langle \varpi_0 \rangle = \langle p_0^2 \rangle_S - \langle p_0 \rangle_{S}^2$ is the initial variance in momentum space which for the free wavepacket remain constant over time. By integrating twice one gets the time evolution of the average width of the wavepacket,
\begin{equation}
\label{spread}
    \langle \Delta q_S \rangle (t) = \sqrt{\Delta q_{0,S}^2 + \dot{\xi}_{0,S} t + \frac{\Delta p_{0,S}^2}{m^2} t^2}
\end{equation}
where the subscript zero means the quantity is evaluated at the initial time $t=0$ (in Schr\"odinger picture). The initial value of the time derivative of the variance satisfies $\langle \dot{\xi}_{0,S} \rangle = \frac{1}{m} \left[- i\hbar + 2 \left(\langle q p \rangle_{0,S} -  \langle q \rangle_{0,S} \langle p \rangle_{0,S} \right) \right]$ which is related with the initial correlation between the position and momentum of the wavepacket. For a Gaussian shaped wavepacket one can show that this derivative term vanishes altogether \cite{messiah}. We now express the standard formula \eqref{spread} in the GRW picture by using the set of equations \eqref{si}-\eqref{sf}. Since the Collapse Model only influences average values dynamically with time, initial values remain unchanged. Defining $\langle \Delta q_{GRW} \rangle = \sqrt{\langle \hat{q}^2 \rangle_{GRW} - \langle \hat{q} \rangle_{GRW}^2}$, and using  \eqref{si}-\eqref{sf} we have the following expansion formula for the width of a free wavepacket in the GRW picture,
\begin{equation}
   \label{spread2}
    \langle \Delta q_{GRW} \rangle (t) = \sqrt{\Delta q_{0,S}^2 + \dot{\zeta}_{0,S} t + \frac{\Delta p_{0,S}^2}{m^2} t^2 + \frac{\alpha \lambda \hbar^2}{6m^2} t^3}.
\end{equation}
We see that the initial values remain unchanged and we get a new term, in addition to the standard Schr\"odinger expansion rate, cubic in time and contributing to an accelerated expansion. Considering once again a Gaussian shape  one has a vanishing derivative term and saturation for the uncertainty principle $\Delta p_0 \Delta q_0 = \hbar/2$. Using this we can express \eqref{spread2} completely in the position space,
\begin{equation}
   \label{gspread}
    \langle \Delta q_{GRW} \rangle (t) = \sqrt{\Delta q_{0,S}^2 +  \frac{\hbar^2}{4 m^2 \Delta q_{0,S}^2 } t^2 + \frac{\alpha \lambda \hbar^2}{6m^2} t^3}.
\end{equation}
The above equation is quite extraordinary -- a single equation bearing the classical, quantum and collapse effects.  Setting $\hbar = 0$ would turn off both the quantum and collapse contributions signalling, once again, that the noise corresponding to the collapse parameter $\lambda$ is purely quantum. On the other hand, by setting $\lambda=0$ we just turn of the collapse effect and thereby end up with conventional quantum result. It is interesting to see that a direct effect of collapses of wavefunctions in fact increases the rate of expansion the wavepacket. This is expected due the ``kicks'' generated by occasional collapses of the wavefunctions constructing the wavepacket itself. To explain clearly, recall that the in standard QM the average velocity the wavepacket is $\langle p \rangle/m$ which provides a one dimensional displacement of the wavepacket. In addition, there is an expansion of the average width. Collapse generated random clicks affects the both. That is,  the center of the wavepacket undergoes an additional yet very mild random displacement (such as in Brownian motion) around its mean positions and every collapse event releases of slight energy (which is  a characteristic feature of collapse models) that in turn makes the wavepacket to undergo an accelerated expansion as represented by the cubic term in \eqref{gspread}.

To compare strength of the collapse mechanism to the standard quantum effect we define a new parameter ``Collapse-to-Quantum Ratio (CQR)''  using \eqref{gspread}. The CQR is a ratio between the collapse term (fourth term inside the squared-root) and the quantum contribution (third term inside the squared-root), defined as
\begin{equation}
\label{cqr}
    \varrho (t;\alpha,\lambda) := \frac{2}{3}(\alpha\lambda \Delta q_{0,S}^2) t.
\end{equation}
Replacing $\alpha$ by the collapse length scale $r_C$ we can express
\begin{equation}
\label{cqr1}
    \varrho (t;r_C,\lambda) := \left(\frac{2\lambda \Delta q_{0,S}^2}{3r_C^2}\right) t.
\end{equation}
To give an estimate consider a characteristic value for the collapse length scale $r_C=10^{-7}$m and for hydrogen $\Delta q_0^2 \simeq 10^{-20}$ m$^2$. Thus for the free Hydrogen molecular wavepacket
\begin{equation}\label{new4}
    \varrho (t;\alpha,\lambda) = \frac{2 \times 10^{-6} \lambda t}{3}.
\end{equation}
Note, however, that original GRW arguments based on macro and micro domains set a very wide bound for the collapse parameter \cite{grw}, $10^{-16}\le \lambda \le 10^{7}$, which has obviously been constrained by various later experiments (as summarized in the Fig. 4 of \cite{space}). Below we show that we are able to not only constrain the collapse parameter $\lambda$ using CQR, but also to provide a  deeper understanding of the classical-quantum division in the collapse parameter space.

\begin{figure}[t!]
\centering
\includegraphics[width=0.45\textwidth]{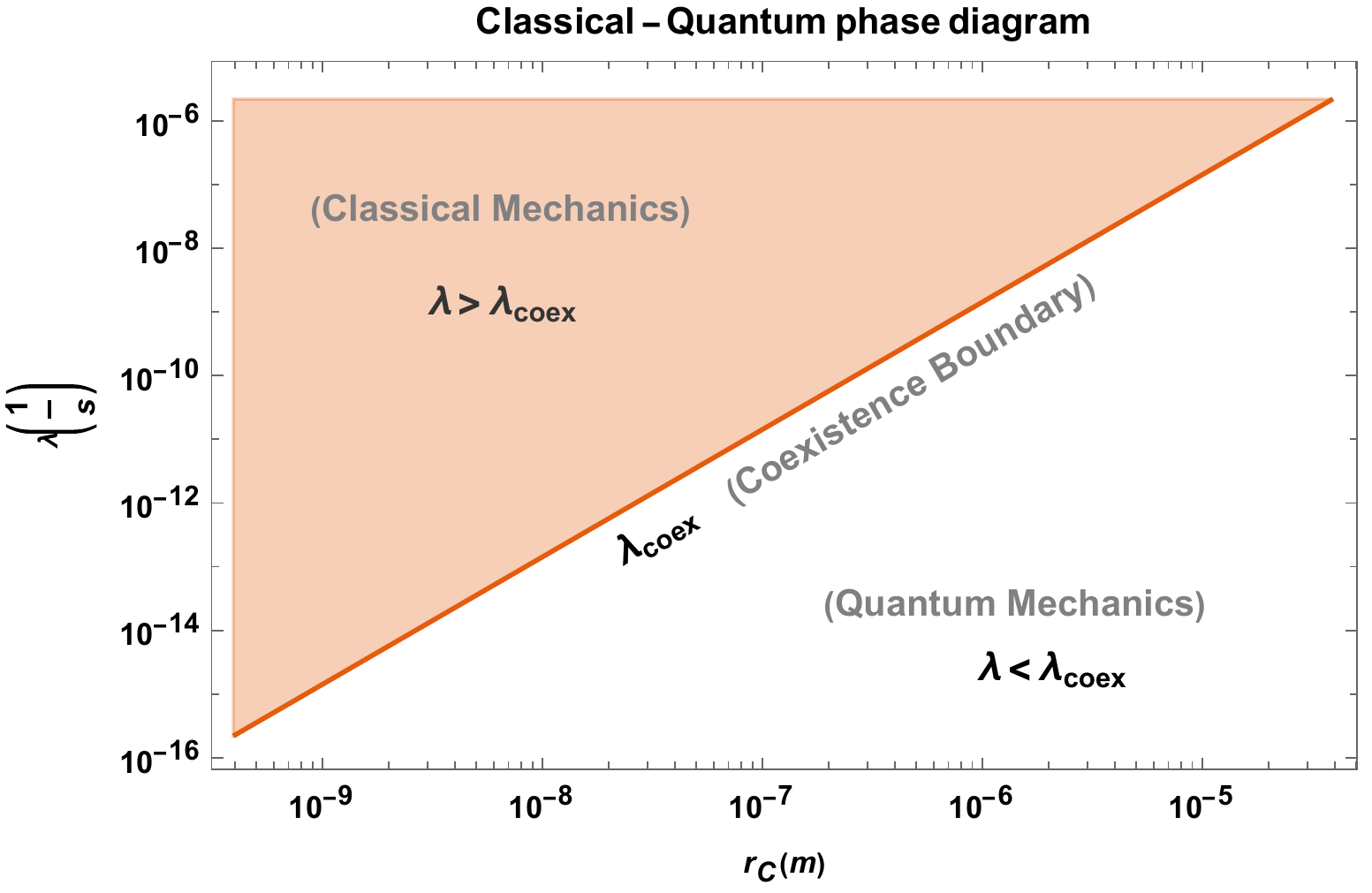}
\caption[]{Unconstrained log-log plot showing the classical and quantum phases for the relevant $\lambda - r_C$ plane. We set $t=10^{17}$, a characteristic value considering the age of the universe in \eqref{cqr}. The plot is closely analogous to the coexistence curve in thermodynamic phase diagram. For details see the text.}
\label{measure-2}
\end{figure}

\begin{figure}[t!]
\centering
\includegraphics[width=0.45\textwidth]{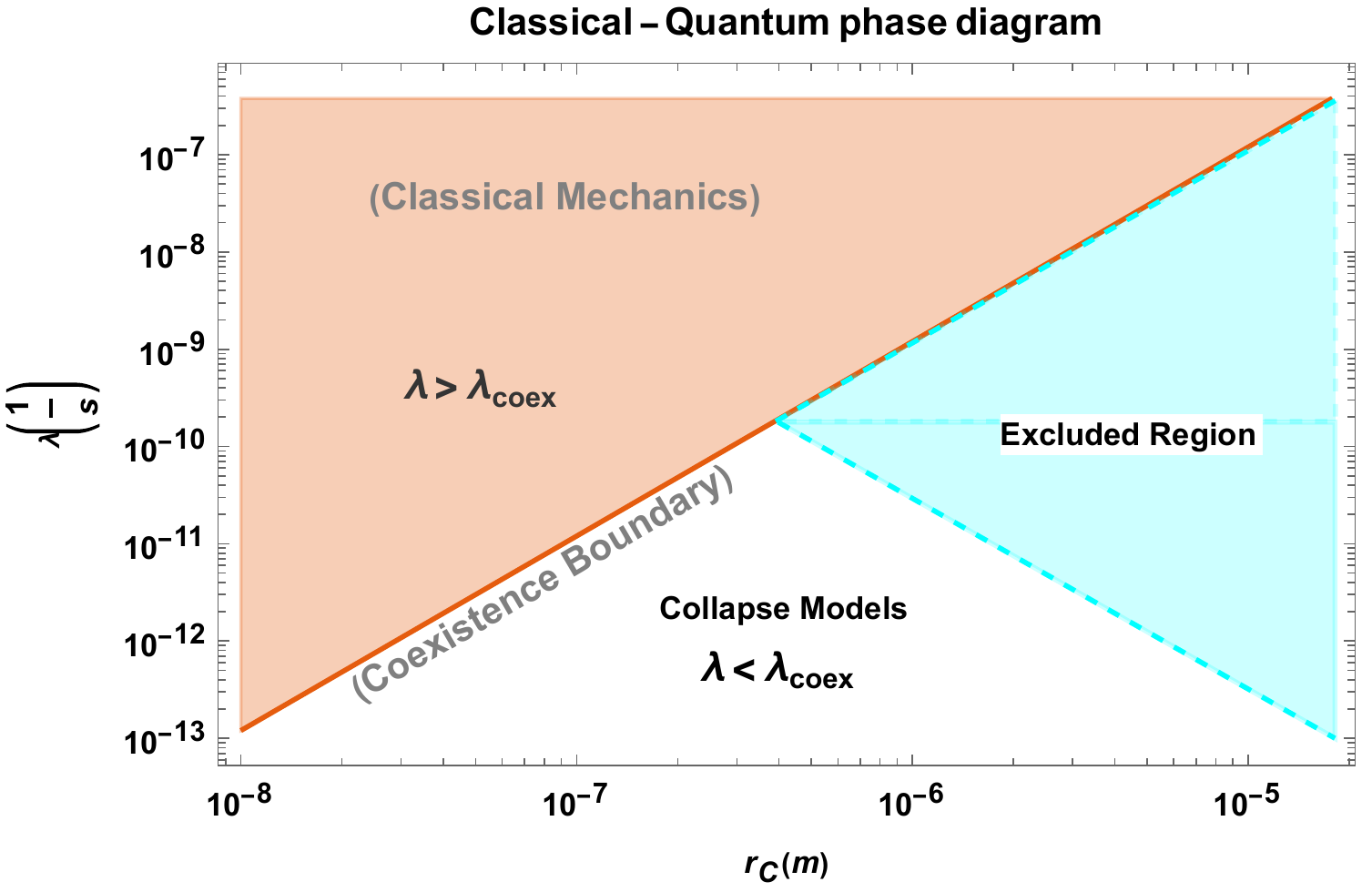}
\caption[]{Log-log plot showing the classical mechanics and Collapse Model parameter spaces in the $\lambda - r_C$ plane for mass-proportional collapse model.  The functional relationship resulting  $\lambda_{\text{coex}}$ line is experimentally supported by the IGEX data \cite{space, igsl}. The parameter space of the phase diagram in  Fig. \ref{measure-2} is reduced after implementing other useful bounds  summarized in Fig. 4 of \cite{space}. For details see the text.}
\label{measure-3}
\end{figure}

The argument for constraining $\lambda$ from our discussion of the free wavepacket dynamics is quite simple:  hydrogen is abundantly found in the universe and has been the main source of energy of stars for billions of years. This stability for a free Hydrogen atom is possible only if the collapse term is strongly subdominant (to the standard quantum mechanical term) even for a timescale comparable to the age of the universe. That is, for ($t\sim 10^{17}$s), we can safely assume that CQR is below unity for Hydrogen and using \eqref{new4} we get, 
\begin{eqnarray}
  \varrho (t=10^{17};\alpha=10^{10},\lambda) &\le & 1\nonumber
  \implies \lambda \le 1.5 \times 10^{-11}.
\end{eqnarray}
Thus from a very general consideration, we can constrain the collapse parameter to an unprecedented narrow band given by  $10^{-16}\le \lambda \le  10^{-11}$. Now given the fact that the collapse length scale $r_C$ can vary from the mesoscopic to near microscopic values, there exist a range of possible values for $r_C$ given by $10^{-8} m\le r_C \le 10^{-4.5} m$. By virtue of \eqref{cqr1} we can generate corresponding bounds on $\lambda$ by using the fact that $\varrho\le 1$. 

At a  deeper level, the very definition of CQR in \eqref{cqr1} also provides a guiding principle for distinguishing classical and quantum systems. Since CQR tracks the strength of collapse dynamics in comparison with the standard quantum dynamics, we can now interpret a system ``strongly quantum'' if $\varrho << 1$ while a system is ``strongly classical'' if $\varrho >> 1$ (i.e, quantum superpositions are strongly suppressed). Therefore, $\varrho=1$ will provide a boundary curve, very similar to a coexistence curve in the phase diagram of thermodynamic systems where both classical and quantum behaviors will show up. For our consideration of Hydrogen atoms originated in the early universe, $\varrho_{\text{coex}}=1$ in \eqref{cqr1} provides a curve $\lambda_{\text{coex}}= (1.5\times 10^3)r_C^2$ which is plotted in Fig. \ref{measure-2} for the allowed range of $r_C$ mentioned before. If experiments show  a continuous weakening of quantum superpositions while crossing the (coexistence) boundary curve in Fig. \ref{measure-2}, one might interpret this ``emeregence of classicality'' or vice-versa as a continuous transition  in thermodynamic phases. In such a scenario one would expect to define a ``quasi-classical'' or ``quasi-quantum'' region around the coexistence curve to be located  by the experimental data. On the other hand, if the breakdown of superpositions is sudden about the coexistence curve in Fig. \ref{measure-2}, one might expect a transition between quantum and classical behaviors which is highly analogous to the first order phase transition in thermodynamic systems.  
It is to be noted  that the analogy of the quantum-classical transition with the phase transition was also proposed earlier \cite{vW1}-\cite{vW4} where breakdown of a global symmetry and emergence of an order parameter in condensed matter systems were related with the non-unitary behavior in quantum dynamics, particularly the collapse of wavefunction. These studies also contributed to a slightly different interpretation of the collapse models \cite{vW3}. In this work, we reach a similar conclusion by staying strictly within the GRW model and from a perspective which is complementary to the studies of \cite{vW1}-\cite{vW3}.
 A strong condition on the boundary curve is in order: for GRW to be good theory, once this curve is defined, such as using Hydrogen atom here, it should not be altered by replacing Hydrogen to some other molecule, just like the value of $\hbar$ does not change by changing the source of radiation itself.

Now we compare our theoretical estimates with the experimental data. A nice summary of various bounds is plotted in $\lambda-r_C$ plane in \cite{space}. They come from variety of experiments started with the LISA Pathfinder data \cite{beyond, non}, Cantilever based experiments \cite{improved}, ultracold layered forced sensors \cite{pearle2}, Gravitational-Wave detectors \cite{grav}, assuming effective collapse rate at the mesoscopic scale \cite{meso} and   the data from International Germanium experiment (IGEX) \cite{miley, bdecay, 95, igsl}.  Although some of these experiments used so called the continuous extension of GRW model (CSL model), the parameter space for $(r_C,\lambda)$ remains very close as in the discrete GRW, and therefore for all practical purposes we can  compare our bound with the results summarized \cite{space}.

Using the definition of CQR in \eqref{cqr}  one derives $\lambda \propto r_C^2$ for an arbitrary but fixed time $t$. This functional dependence, plotted as the coexisting boundary $\lambda_{\text{coex}}$ in Fig. \ref{measure-2}, is extraordinarily supported by the X-ray emission data collected by the IGEX experiment \cite{miley, bdecay, 95, igsl} (summarized in blue dotted line of Fig. 4 in \cite{space}). This match can also be made numerically accurate with one arbitrariness -- one can match both the mass-proportional and non-mass-proportional collapse models considered in \cite{igsl} by suitably adjusting the time $t$ when the CQR $\varrho \ge 1$. 
For the mass-proportional collapse model and considering a Gaussian distribution for the data Piscicchia {\it et al}. \cite{igsl} estimate $\lambda_{m.p} \le 8.1 \times 10^{-12}$ (for $r_C=10^{-7}$m). Here, in Fig. \ref{measure-2}, we can easily translate the bound on $\lambda$ arising from \eqref{cqr} with that in the mass-proportional collapse model \cite{Bassi:2012bg} simply by using the relationship $\lambda_{m.p.} = \lambda \left(\frac{ m_{H}}{{m_n}}\right) = 2\lambda$ where $ m_{H}$ is the mass of a Hydrogen atom and $m_n$ of a nucleon.  Replacing $\lambda$ by $\lambda_{m.p}$ in \eqref{cqr} we can reproduce the bound $\lambda_{m.p} \le 8.1 \times 10^{-12}$ by setting $t\ge 3.70\times 10^{17}$s, which is of the order of the age of the universe. This tells us mass-proportional collapse models cannot be discarded as yet. On the other hand we can also reproduce the non-mass-proportional bound of \cite{igsl} $2.4 \times 10^{-18}$ by using our original definition of CQR in \eqref{cqr} and by setting $t\ge 6.25\times 10^{23}$. Irrespective of the numerical values, physically these inequalities give us an estimate as to when the collapse dynamics start dominating the quantum dynamics on a free particle. In mass proportional model this time is in the hindsight while for non-mass proportional model this time is still 6 orders of magnitude away. This arbitrariness cannot be fixed within our framework, but if experimental evidence prefers one collapse model over the other, then we know exactly when collapse effect will dominate over the quantum effect for a free particle. If we chose the mass-proportional model and other bounds exhibited in Fig. 4 of \cite{space} we can further constrain the Classical-Quantum phase diagram in Fig. \eqref{measure-2} which is represented in Fig. \eqref{measure-3}. It is our opinion that we can simply call all plots in the $\lambda - r_C$ plane (such as Fig. 4 of \cite{space}) as ``classical-quantum phase diagram'' in view of our results. It will be interesting to see how far the proposed analogy between the quantum-classical behavior and that of the phase transitions can be pushed for.

Although we leave the above issue as an open problem and a topic for future research we want to add the following comment on this:  in thermodynamics we have given prescriptions to treat phase transitions and the most closer prescription, considering  the ``classical-quantum'' phase diagrams we propose here, would be using the Clausius-Clapeyron or Ehrenfest type equations to classify this phase transition between the quantum and classical behavior. Just like a $PVT$ system we would need to define parameters analogous to pressure ($P$), volume ($V$) and temperature ($T$). One way to do this is to consider the mass proportional collapse models where we have three similar variables, such as the collapse parameter ($\lambda$), collapse length scale ($r_C$) and the mass ($m$).  It might be possible to construct an analogous $\lambda m r_{C}$ system which just like a $PVT$ would need experimental data on the behavior of the collapse parameter  near the coexistence curve as we change continuous variables such as the mass and size of the system. In the water-vapour system volume (or density) changes abruptly as one crosses the coexistence curve. It would be interesting to see if similar behavior is registered for the collapse parameter. In addition, it will be interesting to experimentally locate the analogous of the critical and triple points if they exist at all. To conclude, reinterpretation of the so called exclusion plot for collapse models as a phase diagram opens up possibilities for further research and we certainly aim to  study these issues in future.

\begin{acknowledgements} 
Authors thank Paolo Amore and Saurya Das for their valuable feedback on the pre-print. Research of FT and SKM are supported by SEP-CONACyT research grant CB/2017-18/A1S-33440, Mexico. 
\end{acknowledgements}

%%%%%%%%%%%%%%%%%

\end{document}